\pdfoutput=1

\documentclass[aps,prb,reprint,showpacs,superscriptaddress]{revtex4-1}

\usepackage{graphicx}
\usepackage{graphics}
\usepackage{amsmath}
\usepackage{amssymb}
\usepackage{amsfonts}
\usepackage{dcolumn}
\usepackage{dsfont}
\usepackage{latexsym}
\usepackage{rotating}
\usepackage{color}
\usepackage{latexsym}
\usepackage{bbm}
\usepackage{subfigure}
\usepackage{float}
\usepackage{epsfig}
\usepackage{epsf}
\usepackage{psfrag}
\usepackage{bm}
\usepackage{amsthm}
\usepackage{eucal}
\usepackage{mathrsfs}
\usepackage{url}
\usepackage{braket}

\usepackage{color} 

\usepackage{hyperref}
\hypersetup{
colorlinks=true,final=true,
        linkcolor=blue,
        citecolor=blue,
        filecolor=blue,
        urlcolor=blue,
}

\newcommand{\femn}{Fe$_{1-x}$Mn$_x$V$_2$O$_4\,\,$}

\newcommand{\phrl}[1]{Phys.~Rev.~Lett. {\bf #1}}
\newcommand{\phrb}[1]{Phys.~Rev.~B {\bf #1}}
\newcommand{\bib}{\bibitem}

\begin{document}

\title{ Orbital Ordering in \femn: A First Principles Study}

\author{Dibyendu Dey}
\email{dibyendu@phy.iitkgp.ernet.in}
\affiliation{Department of Physics, Indian Institute of Technology Kharagpur, Kharagpur - 721302, India}
\author{T. Maitra}
\affiliation{Department of Physics, Indian Institute of Technology Roorkee, Roorkee - 247667, India}
\author{A. Taraphder}
\affiliation{Department of Physics, Indian Institute of Technology Kharagpur, Kharagpur - 721302, India}
\affiliation{Centre for Theoretical Studies, Indian Institute of Technology Kharagpur, Kharagpur - 721302, India}

\date{\today}
\begin{abstract}
Long range orbital order has been investigated in \femn as a function of 
doping (x)
using first principles density functional theory calculations
including the effects of Coulomb correlation and spin-orbit interaction within
GGA+U and GGA+U+SO approximations. Through a detailed analysis of corresponding 
Wannier orbital projections of the Vanadium d bands, we have clearly established 
that for x$\le$0.6, the orbital order at V sites 
consists of a linear superposition of d$_{xz}$ and d$_{yz}$ orbitals of the 
type d$_{xz}\pm$d$_{yz}$. Within each ab-plane a ferro-orbital ordering of 
either d$_{xz}$+d$_{yz}$ or d$_{xz}$-d$_{yz}$ is observed which alternates 
in successive ab-planes along the c-direction. On the contrary, for x$>$0.6, it is 
the d$_{xz}$ or d$_{yz}$ orbital which orders at V sites in successive 
ab-planes along c-direction (so called A-type ordering). At Fe sites, we 
observe an orbital ordering of d$_{x^2-y^2}$ orbitals for x$\le$0.6 
and d$_{z^2}$ orbitals for x$>$0.6. Effect of spin-orbit interaction on 
orbital ordering is found to be not significant in the entire range of doping studied.
\end{abstract}
\maketitle

\section{Introduction}
Correlated and frustrated electronic (spin) systems have recently observed a 
major surge in research activities due to the inherently rich fundamental 
physics involved in their understanding as well as their potential for 
practical applications. Vanadium spinel oxides belong to one such family of 
systems which, of late, has become the centre of attraction for studying phenomena involving 
the competition among various degrees of freedom, like spin, orbital and lattice 
in presence of both Coulomb correlation and geometric frustration \cite{Lee,Rad}.
These compounds with chemical formula AV$_2$O$_4$ can be broadly divided into 
three categories on the basis of A-site cation properties, namely; 
(i) A site non-magnetic (i.e. A= Zn, Mg, Cd), 
(ii) A-site magnetic but orbitally inactive (i.e. A = Mn, Co), 
(iii) A-site magnetic and orbitally active (i.e. A = Fe). 
As V ion is both magnetic and orbitally active, the additional magnetic and orbital 
degrees of freedom of A site ion bring in a very complex and rich physics with 
competing spin, lattice and orbital interactions (in some cases also spin-orbit 
interactions) in these systems.
The competition among different degrees of freedom then manifests itself in 
several structural (accompanied often by orbital ordering) and magnetic transitions 
as observed experimentally in these systems.    

FeV$_2$O$_4$ is a rare example among the Vanadium spinel oxides where both
Fe and V cations are magnetic as well as orbitally active. This system is 
reported to show three successive structural transitions (cubic$\rightarrow$ 
tetragonal$ \rightarrow$ orthorhombic $\rightarrow$ tetragonal) and two 
magnetic transitions (paramagnetic (PM) $\rightarrow$ collinear ferrimagnetic 
(FIM) $\rightarrow$ non-collinear ferrimagnetic (NC-FIM)) as the 
temperature is lowered \cite{Katsu,Nii1,MacD,Kang}.
Another example is MnV$_2$O$_4$ which has A-site magnetic but not orbitally 
active. This compound shows one structural (cubic$\rightarrow$ tetragonal) and 
two magnetic ( PM $\rightarrow$ FIM $\rightarrow$ NC-FIM) transitions 
\cite{Nii1,Suzuki,Garlea,Tsunet,Adachi}.
ZnV$_2$O$_4$ can be the third example with a nonmagnetic and non-orbitally 
active A-site ion which undergoes one structural (cubic$\rightarrow$ 
tetragonal) and one magnetic (PM $\rightarrow$ antiferromagnetic) transition \cite{Ree}.
Although most of the structural and magnetic transitions (along with the 
magnetic orders) have been unambiguously established by various experiments, 
the orbital order which often accompanies the structural transitions are 
still a matter of intense investigation and continuous debate \cite{Nii1,cherny,Motome,TM2,TM1,TSD}.

The debate is mostly centered around the orbital order of the V 
sites in these spinels. V ion in its ${3+}$ valence state has two
$d$-electrons in the orbitally degenerate $t_{2g}$ levels. In most of these systems, 
low temperature structure is tetragonal and VO$_6$ octahedra
have tetragonal compression accompanied often by a trigonal distortion. 
Due to the tetragonal compression along z-direction the energy of the 
d$_{xy}$ orbital is lowered and it is occupied by one d electron. However, 
the second d electron has a choice: occupy either d$_{xz}$ or d$_{yz}$ orbital or a 
combination of these two. There are various possible 
scenarios proposed for the orbital order in these systems based on
the symmetry of the tetragonal phase and the type of dominant interactions. 
Some of the proposed orbital ordering models include ferro-orbital ordering of
complex orbitals such as d$_{xz}$$\pm$id$_{yz}$ with dominant spin-orbit 
interaction\cite{TM2,cherny}, A-type antiferro-orbital ordering of real orbitals d$_{xz}$ and d$_{yz}$ 
alternating along c-direction where SO interaction is not considered strong\cite{Garlea,Tsunet,Motome} 
and an orbital ordering of real and complex orbitals for systems with intermediate 
strength of SO interaction\cite{Wheeler,Raman} etc. Also, experiments have not yet been able to establish
the orbital order unambiguously in these systems as there are  
measurements reported in the literature supporting each of 
these models. 

Very recently, the doped variants of FeV$_2$O$_4$ and MnV$_2$O$_4$  
(i.e. \femn ($0\le x \le 1$)) have been studied experimentally by Kawaguchi et 
al.\cite{Kawa} and D. Choudhuri et al.\cite{DC} where a detail magnetic and structural phase diagram has been obtained. 
Kawaguchi et al. have also analyzed the orbital ordering at Fe and V sites in \femn for different 
doping (x) on the basis of local Jahn-Teller distortions of FeO$_4$ tetrahedra and VO$_6$ 
octahedra respectively. From their analysis, the authors have predicted that in the 
range 0$\le x \leq 0.6$, the d$_{x^2-y^2}$ orbital should order at all Fe 
sites and at V sites a long range ferro-orbital ordering of complex orbitals 
of the type d$_{xz}$$\pm$id$_{yz}$ should be present. On the other hand, for the range 0.6 $< x \leq 1$, it is 
the d$_{z^2}$ orbital that orders at Fe sites and an A-type antiferro-orbital 
order should exist at V sites where within each ab-plane it is ferro-orbital
order of either d$_{xz}$ or d$_{yz}$ but along c-direction it is
antiferro (i.e. alternate arrangement of d$_{xz}$ and d$_{yz}$ orbital).   
However, a very recent XMCD measurements reveal that though there is a 
difference in the occupied orbital state in FeV$_2$O$_4$ and MnV$_2$O$_4$, the 
orbitals that order (at V sites) in FeV$_2$O$_4$ as well as in MnV$_2$O$_4$ are not complex 
but real orbitals \cite{Nii2}. Even though the observation of orbital order 
consisting
of real orbitals rather than complex ones corroborates the previous theoretical predictions reported in the literature on these two systems\cite{TM1,TSD}, the exact nature
of orbital ordering proposed from theory and experiments differ significantly. 
In view of these disagreements regarding the orbital state of parent compounds and also to study the effect of doping, we have decided to investigate the issue of orbital order in 
the whole range of x in Fe$_{1-x}$Mn$_x$V$_2$O$_4$ from a theoretical perspective 
using ab-initio density functional theory calculations.    
 
\section{Methodology}
We have performed ab-initio density functional theory calculations within 
various approximations such as GGA, GGA+U and GGA+U+SO for \femn with x=0, 0.25,
0.5, 0.75 and 1. The experimental structures of parent and doped compounds have 
been taken from experiments \cite{Nii1,Kawa,DC}. In 
case of x=0.25 and x=0.75 we have taken the available experimental structural 
data for x=0.2 and x=0.8 and optimized the structure as described below.
The optimization (volume, c/a ratio and oxygen positions) has been done by 
using Perdew-Burke-Ernzerof Generalized Gradient Approximation (PBE-GGA) exchange-correlation functional
\cite{Perdew}, within the full potential 
linearized augmented plane-wave (FP-LAPW) method as implemented in the WIEN2k 
code \cite{Wien}. We have used the package 2DRoptimize available within WIEN2k
package for the optimization of volume and c/a ratio. 
For this purpose we have considered nine different volumes around the experimental volume and 
for each volume nine c/a ratios were considered. During the optimization
involving these 81 structures, the free internal parameters (here oxygen 
positions) were also optimized. Final optimized structural parameters were obtained
by the analysis of the following data; 
(i) Energy vs. c/a for each volume, 
(ii) energy vs. volume (with optimized c/a) and 
(iii) c/a vs. volume. 
Each curve was then fit to 3rd, 4th and 5th order of polynomial and the 
lattice constants a and c were calculated from the best fit. We have listed
some optimized structural parameters for x = 0, 0.5 and 0.75 in Table I which we have used
 for our calculations reported here. We have considered the muffin-tin radii 
of 2.00, 1.95, 1.65 a.u for Fe(Mn), V and O respectively. 
The plane wave cutoff RK$_{max}$ was set to 8.0 for all calculations and the 
number of $\vec{k}$ points in the irreducible wedge of the Brillouin zone (BZ) 
were taken to be 60 for x = 0, 0.25, 0.5 and 72 for x = 0.75, 1. The expansion of the radial wavefunction in spherical harmonics was 
considered up to angular momentum quantum number $l = 10$. We also performed 
calculations  
including the correlation effects arising from the d electrons of Fe/Mn and V.  
For this, we have used a self-interaction corrected (SIC) GGA+U \cite {Ani} approximation which takes 
into account the on-site Coulomb interaction U and removes the self-Coulomb and self exchange correlation energy. 
Spin-orbit (SO) coupling is included by the second variational method \cite {Koe} with scalar relativistic wavefunctions. 
All the calculations were performed for the collinear ferrimagnetic spin arrangement of 
Fe/Mn and V ions.
WANNIER90 \cite{Wannier} and WIEN2WANNIER \cite{w2w} (the interface program 
between WIEN2k and WANNIER90) were used to fit the bands around Fermi level
by localized Wannier orbitals and to analyze the band characters in terms of
these orbitals.

\section{Results and Discussion}

On the basis of symmetry of the crystal structure and observed orbital 
ordering in our calculations, we divide these compounds into two groups 
(i) x=0, 0.25, 0.5 which we refer to as x$\le$0.6 case as crystal symmetry
remains same in the range 0$\le$x$\le$0.6, (ii) x=0.75, 1 which we refer
to as x$>0.6$ case where symmetry is different from the former one.

\subsection{\femn; x=0, 0.25, 0.5}
\begin{figure}
\begin{center}
\includegraphics[width=7.5cm]{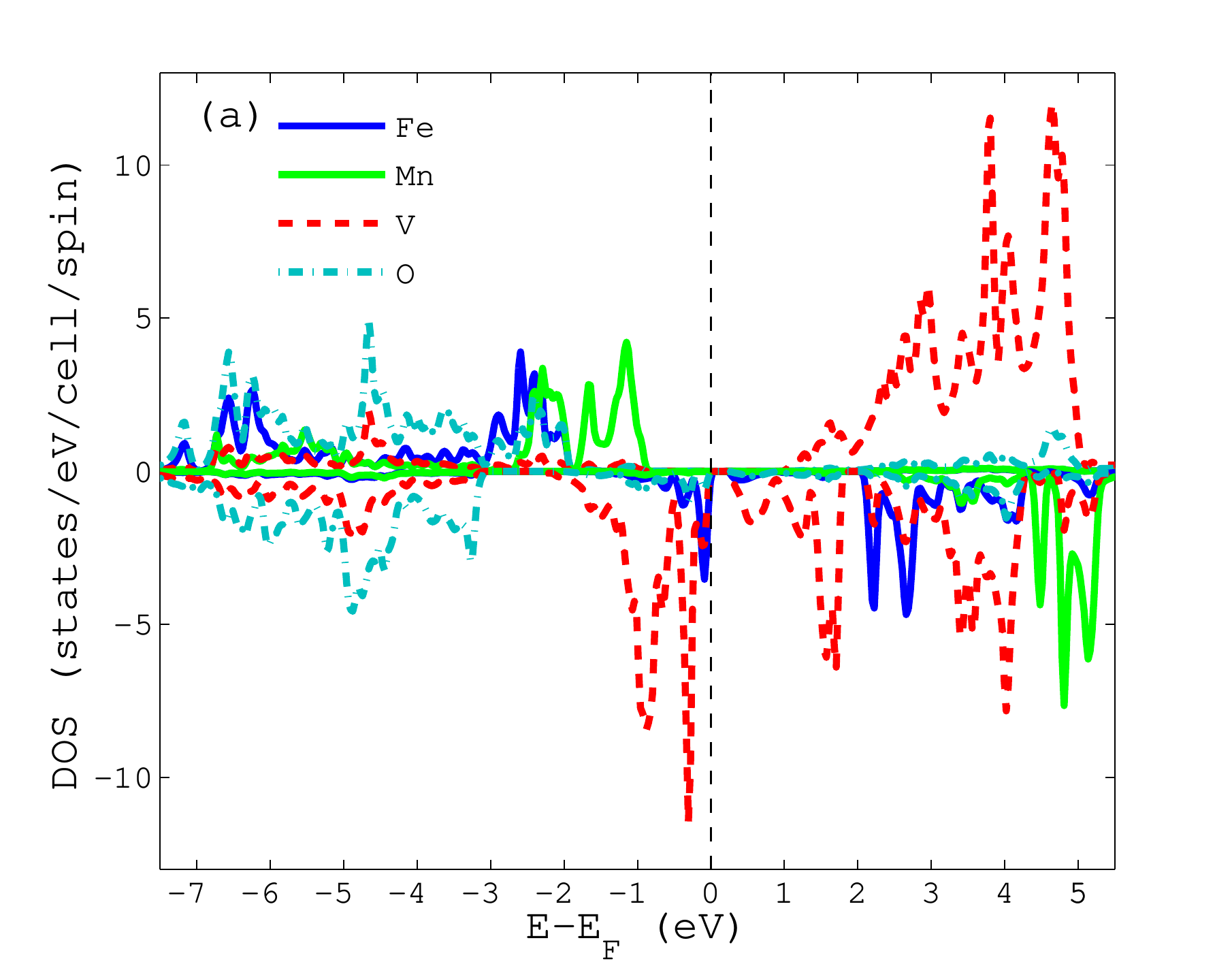}
\includegraphics[width=7.5cm]{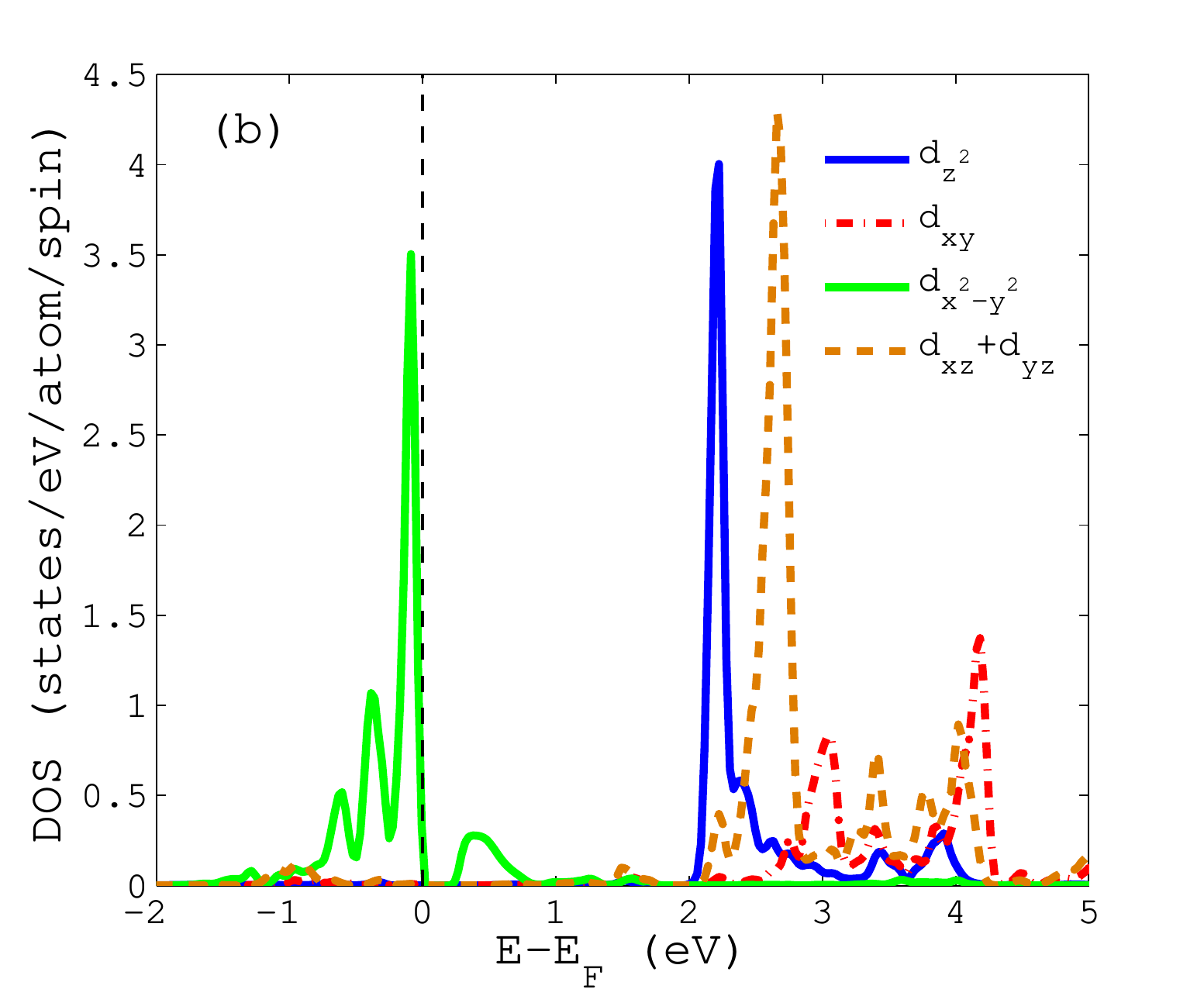}
\includegraphics[width=7.5cm]{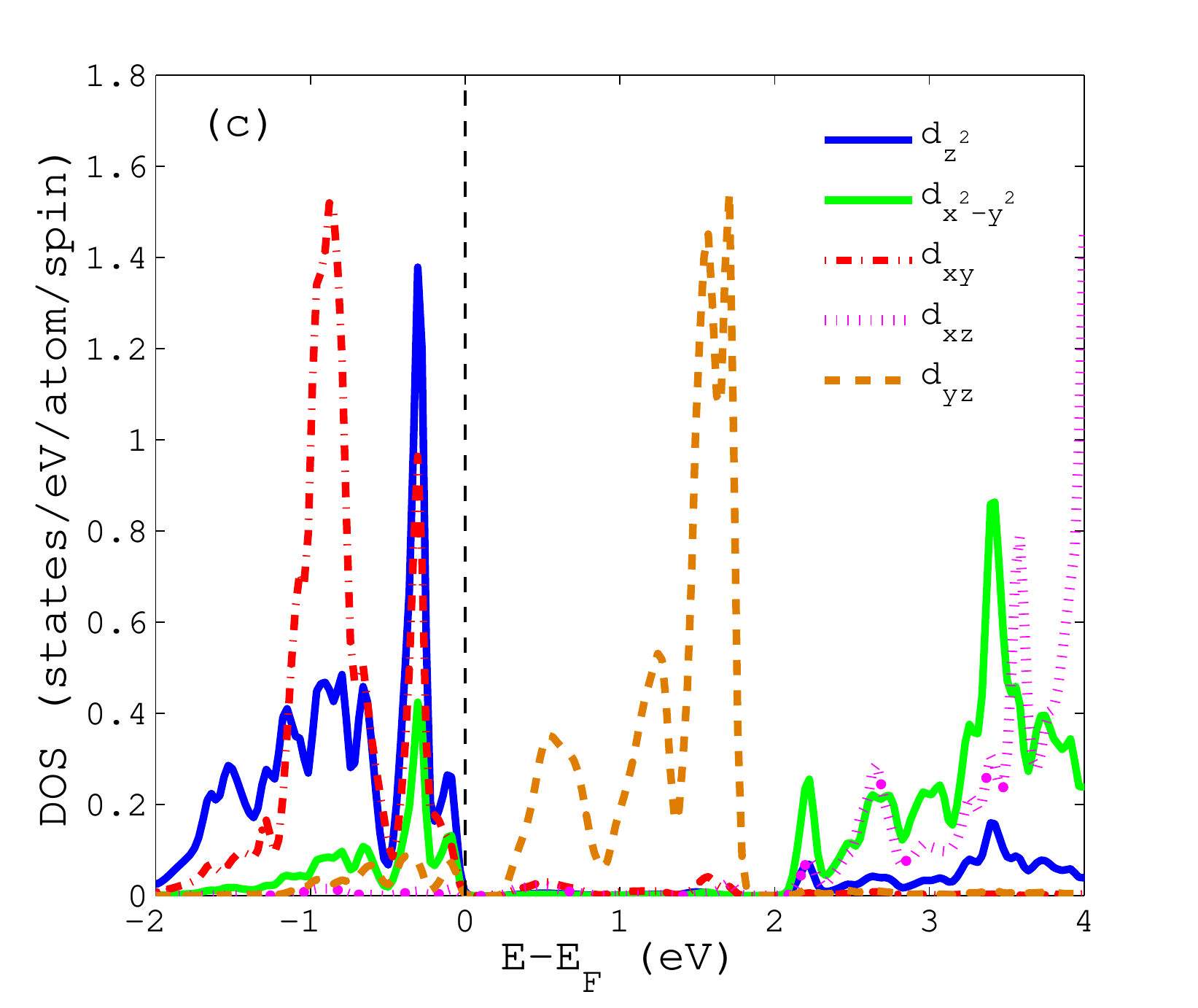}
\caption{\label{dos_50} (Colour online) Spin polarized DOS are shown for x=0.5. (a) Total DOS, (b) partial DOS of Fe in minority spin channel (c) partial DOS of V in majority spin channel.}
\end{center}
\end{figure}

In the low temperature tetragonal phase, the crystal structures of \femn at 
x=0, 0.25, 0.5 have the I4$_1$/amd symmetry as observed experimentally
\cite{Nii1,Kawa,DC}. Our optimized structures also have the same symmetry 
although the local trigonal and tetragonal distortions as well as the c/a
ratios differ slightly. In Table I we compared the structural parameters
of experiment and optimized one for x=0.5 case. We clearly observe from Table I
that the local trigonal and tetragonal distortions at FeO$_4$ tetrahedron 
and VO$_6$ octahedron as well as the c/a ratio of the optimized structure are comparable and more
importantly have the same signs (i.e. compression or elongation along a particular direction) 
as the experimental one. Therefore, we discuss below the results obtained with these optimized structures. 
We note that 
we have also performed calculations with the corresponding experimental 
structures where it is available and checked that our both experimental and
optimized structures give consistent results. 

\begin{figure}
\begin{center}
\includegraphics[width=9cm]{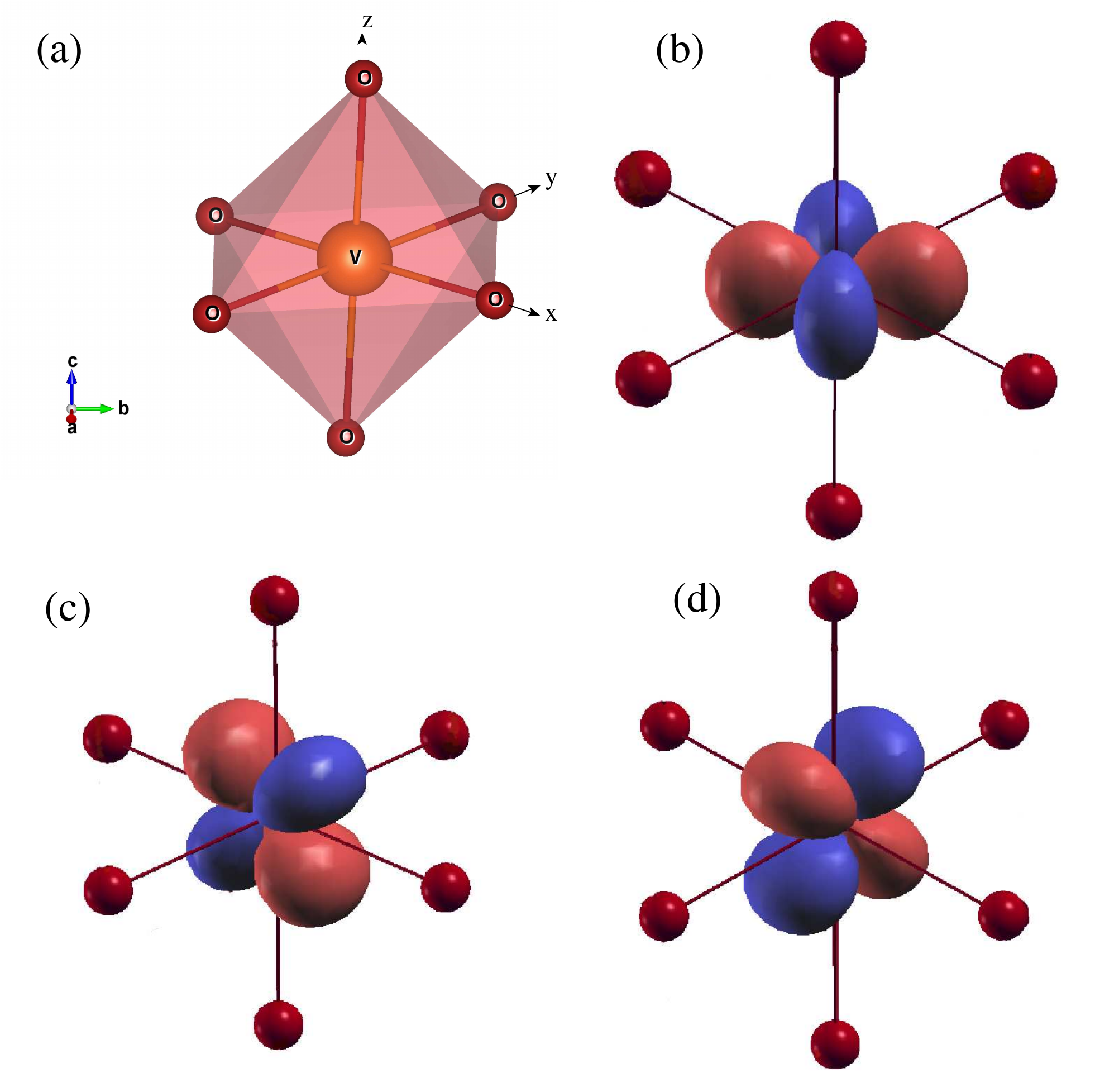}
\caption{\label{wan_orb} (Colour online) Orientation VO$_6$ octahedron is shown in (a) with 
respect to reference frame of the tetragonal unit cell (abc)  and also shown
is the xyz frame in which Wannier orbitals are defined. The d$_{xy}$, d$_{xz}$ 
and d$_{yz}$ Wannier orbitals are shown in (b), (c) and (d) respectively.}
\end{center}
\end{figure}

\begin{table}
\centering
\begin{tabular}{|p{2.0cm}|p{1.5cm}|p{1.5cm}|p{1.5cm}|p{1.5cm}|}
 \hline
   & Lattice Constants & O-Fe-O & V-O & O-V-O \\
 \hline
 \hline
 Optimized; x=0 & a=5.9530, c=8.4485, c/a=1.004 & 107.078$^{\circ}$ (along-c) 110.681$^{\circ}$ & 1.9933 A$^{\circ}$ (along-c) 2.0351 A$^{\circ}$ & 85.597$^{\circ}$ 94.403$^{\circ}$\\
  \hline
 Experimental; x=0.5 & a=5.9863, c=8.4994, c/a=1.004  & 108.004$^{\circ}$ (along-c) 110.210$^{\circ}$  & 2.0079A$^{\circ}$ (along-c) 2.0283A$^{\circ}$ &  84.493$^{\circ}$ 95.507$^{\circ}$\\
 \hline
 Optimized; x=0.5 & a=5.9497, c=8.4497, c/a=1.004 & 108.649$^{\circ}$ (along-c) 109.884$^{\circ}$ & 2.0318A$^{\circ}$ (along-c) 2.0361A$^{\circ}$ & 85.612$^{\circ}$ 94.388$^{\circ}$\\
 \hline
 Optimized; x=0.75 & a=6.0502, c=8.40, c/a=0.982 & 111.984$^{\circ}$ (along-c) 109.884$^{\circ}$ & 2.03A$^{\circ}$ (along-c) 2.032A$^{\circ}$ 2.035A$^{\circ}$ & 83.715$^{\circ}$ 96.285$^{\circ}$\\
 \hline
\end{tabular}
\caption{Structural parameters and various distortions observed in the 
optimized structures for x=0, x=0.5 and x=0.75 are listed. Corresponding experimental data is also given in case of x=0.5 case for comparison\cite{DC}.}
\label{Tab:1}
\end{table}
      
{\bf GGA+U Results:} As both parent compounds (FeV$_2$O$_4$ and MnV$_2$O$_4$) as well as the doped 
ones are known to be Mott insulators, we have included in our calculations the Coulomb
correlation within GGA+U approximation. The calculations have been performed for various 
values of U$_{eff}$ ($=U-J$, where U is on-site Coulomb interaction and J is the Hund's exchange 
interaction strength) ranging from 3eV to 5eV  at Fe, Mn and V sites. 
We note that conclusions drawn from our calculations do not change when we vary the values
of U$_{eff}$ in the above range. Therefore, we present below the results obtained for the 
U$_{eff}$ value of 4eV (3eV) at Fe/Mn (V) sites. In addition, in all our calculations we 
have considered collinear ferrimagnetic order among the Fe (Mn) and V spins (i.e. V spin 
moments are aligned opposite to Fe (Mn)
spin moments).

\begin{figure}
\vspace{-1.0cm}
\begin{center}
\includegraphics[width=9.2cm]{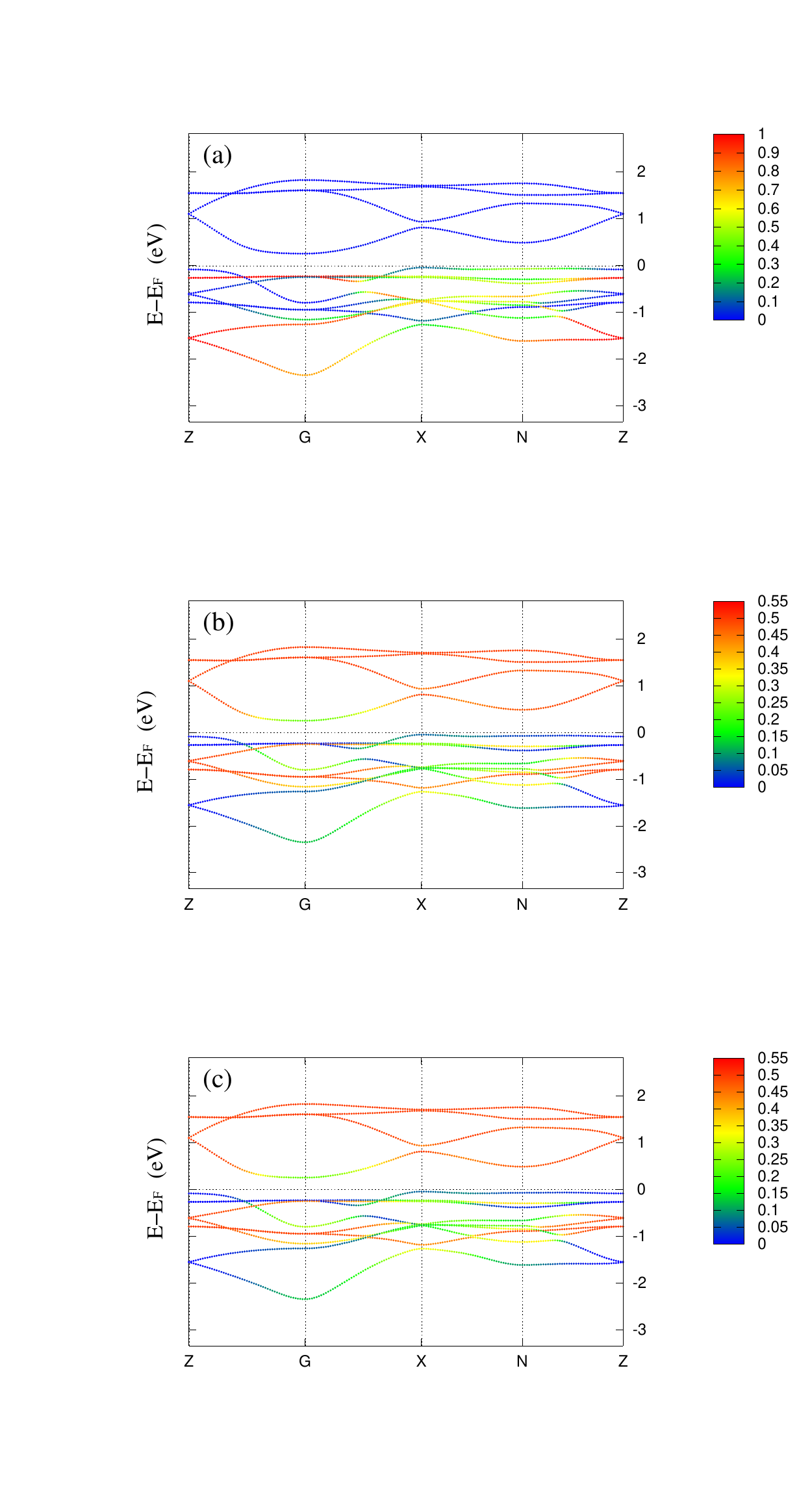}
\caption{\label{wan_proj_50} (Colour online) Projections of V t$_{2g}$ bands on to d$_{xy}$, d$_{xz}$ and d$_{yz}$ Wannier orbitals are shown for x=0.5 in (a), (b) and (c) respectively. It is seen that d$_{xy}$ has the lowest energy and it is fully occupied while
d$_{xz}$ and d$_{yz}$ are occupied with equal weight.}
\end{center}
\end{figure}

Our primary goal here is to ascertain the type of ground state orbital order
in these systems. As mentioned above we observed qualitatively similar 
behaviour as far as the orbital order is concerned for x=0, 0.25 and 0.5. Therefore,  we present the case of x = 0.5 in detail below. In Fig. 1(a) we show 
the total density of states (DOS) of Fe, Mn, V and O in
both the spin channels within GGA+U approximation. Consistent with the experimental observation the insulating ground state with a gap of about 0.2eV is 
obtained. Furthermore, we see that due to the tetrahedral crystal field, the 
Fe d- states are split into $t_{2g}$ and $e_g$ with $e_g$ states having lower 
energy. The valence state of Fe is $2+$ as the spin up (majority) states are 
fully filled by 5 d electrons and spin down (minority) $e_g$ states are
partially filled by one electron. Partial filling of the minority $e_g$ states 
causes Jahn-Teller effect which in turn further splits these states to
lower energy d$_{x^2-y^2}$ and higher energy d$_{z^2}$ state. This 
energy level splitting can be understood by looking at the local distortions 
of FeO$_4$ 
tetrahedron presented in Table I. In the high temperature cubic phase, all six 
interior O-Fe-O bond angles are degenerate
having a value 109.4$^o$. Whereas, in the low temperature tetragonal phase 
FeO$_4$ tetrahedron is elongated along
c-direction making d$_{x^2-y^2}$ the ground 
state orbital. Therefore, the minority spin 
electron occupies d$_{x^2-y^2}$ orbital as we can see in the partial DOS of Fe 
d-states in the minority spin channel presented in Fig 1(b).   
  
Similarly, Vanadium d-states are split into $t_{2g}$ and $e_g$ with $t_{2g}$ 
states having the lower energy due to octahedral crystal field. Majority (here spin 
down due to ferrimagnetic ordering) $t_{2g}$ states are partially filled by 
two d electrons. Unlike FeO$_4$ tetrahedron (where all Fe-O bond lengths are equal), 
VO$_6$ octahedron has both tetragonal and trigonal distortions in the low temperature 
tetragonal phase (see Table I). There is a compression of 
V-O bonds along c-direction compared to four equal V-O bonds in the ab-plane.  
In addition, there also exists a trigonal distortion (i.e. O-V-O bond angles
deviating from 90$^o$) in the ab-plane as well as out of ab-plane. 
Because of the presence of tetragonal compression along c(z) direction the
degeneracy of the $t_{2g}$ states is further lifted and d$_{xy}$ band moves to
 a lower energy below Fermi level (FL) and gets fully occupied by one electron. One can clearly observe this from the partial DOS of V d-states (in majority spin channel) presented
in Fig 1(c). We also observe from Fig 1(c) that there is a further splitting of
the remaining two $t_{2g}$ states to accommodate the second electron giving rise to the insulating gap at the FL.
The later splitting of $t_{2g}$ states happens due to the orbital ordering.
In order to find out the orbital character of the second highest occupied state
, we have calculated the projections of three Wannier
orbitals as shown in Fig 2 of the Vanadium d-bands around the FL. The 
corresponding projections are shown in Fig 3(a), (b) and (c) for d$_{xy}$, 
d$_{xz}$ and d$_{yz}$ orbitals respectively. From these figures we can clearly establish 
that d$_{xy}$ orbital character is the most dominant one among the lower energy 
bands and it's peak has a lower energy than the other two. 
It appears only below FL and hence fully occupied which is 
expected because of the compression of V-O bonds along c(z)-direction. Interestingly 
we further observed that the d$_{xz}$ and d$_{yz}$ characters of exactly equal 
proportions are present in higher energy bands (i.e. close to FL) and these are partially filled. 
This implies that d$_{xy}$ orbital has the lowest energy and it is filled by 
one electron but the 2nd electron occupies equally d$_{xz}$ and d$_{yz}$
orbitals (i.e. a linear superposition of d$_{xz}$ and d$_{yz}$ orbitals).

\begin{figure}
\begin{center}
\includegraphics[width=8cm]{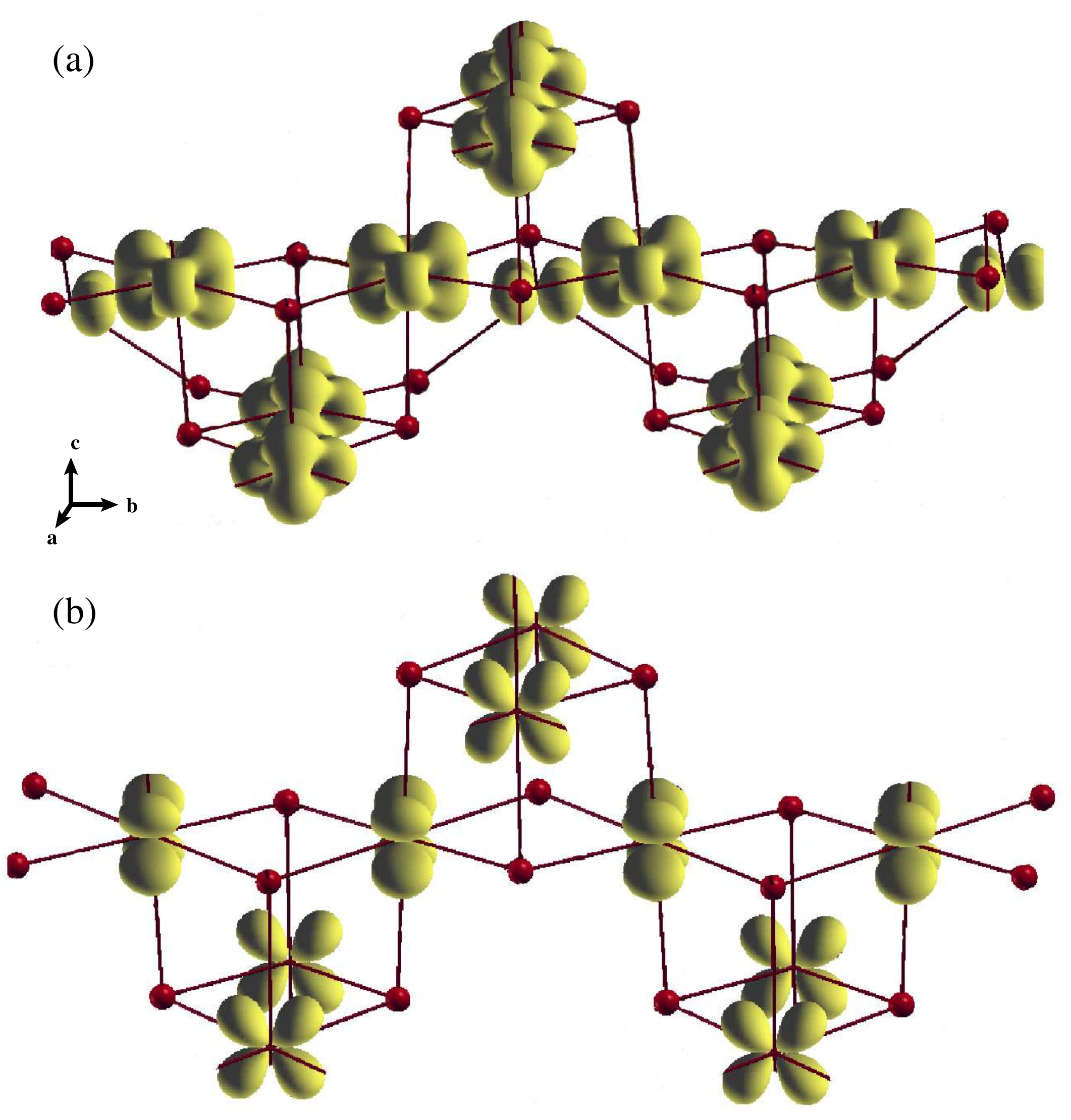}
\caption{\label{el_den_50} (Colour online) 3D electron density is shown in real space lattice
for x=0.5 case in (a). Corresponding hole densities are shown in (b) to depict
the nature of unoccupied orbitals.}
\end{center}
\end{figure}

In order to find out the type of long range orbital order that exists in these
systems and the exact composition of the orbitals that order we calculated 
the 3D electron density in real space and presented it in Fig. 4(a) (for x=0.5). 
As expected we can now clearly see that there is an ordering of d$_{x^2-y^2}$ 
orbitals at all 
Fe sites. The V orbitals ordering however is more complicated. In each ab-plane V orbitals
are ferro-orbitally ordered with the second electron occupying either 
$\frac {1}{\sqrt{2}} (d_{xz}+d_{yz})$ or $\frac {1}{\sqrt{2}} (d_{xz}-d_{yz})$ 
at all V sites. As we move along c-direction, in successive 
ab-planes we observe $\frac {1}{\sqrt{2}} (d_{xz}+d_{yz})$ or $\frac {1}{\sqrt{2}} (d_{xz}-d_{yz})$ alternately ordered. To depict this clearly we have 
also plotted the corresponding hole densities (i.e. for V states just above FL) 
of V 
chains in adjacent ab-planes (see Fig 4(b)). The staggered orbital order along 
c-direction can be clearly seen here. We observe from Fig. 4 that the electron
orbitals point towards each other along a V chain whereas hole orbital point
away from each other which could be understood on the basis of kinetic energy
gain due to more overlap between the neighbouring V orbitals. 

{\bf GGA+U+SO Results:} As it is established from various experimental and theoretical studies that 
some of these Vanadium spinel oxides also show dominant spin-orbit (SO) 
interaction effect which then leads to the orbital ordering of complex orbitals
rather than real ones\cite{TM2,cherny}, we have included SO interaction in our calculations 
through GGA+U+SO approximation. However, in the entire series of compounds studied, the 
effect of SO interaction was not found to be significant either on the 
electronic structure or on the orbital ordering. Orbital moments
also observed to be quite small compared to, for example, ZnV$_2$O$_4$ where 
it is found to be large. We list the calculated orbital moments for Fe and V
for different doping in Table II. Interesting, in a recent XMCD measurement, 
Nii et al. also 
observed negligible orbital moment in both FeV$_2$O$_4$ and 
MnV$_2$O$_4$\cite{Nii2}. Therefore, we conclude that 
the long range orbital order that is present at V and Fe sites as discussed above involves 
only real orbitals not the complex ones such as d$_{xz}$$\pm$id$_{yz}$ which was observed in case of ZnV$_2$O$_4$\cite{cherny,TM2}. 

\subsection{\femn; x=0.75, 1}
\begin{figure}
\begin{center}
\includegraphics[width=8cm]{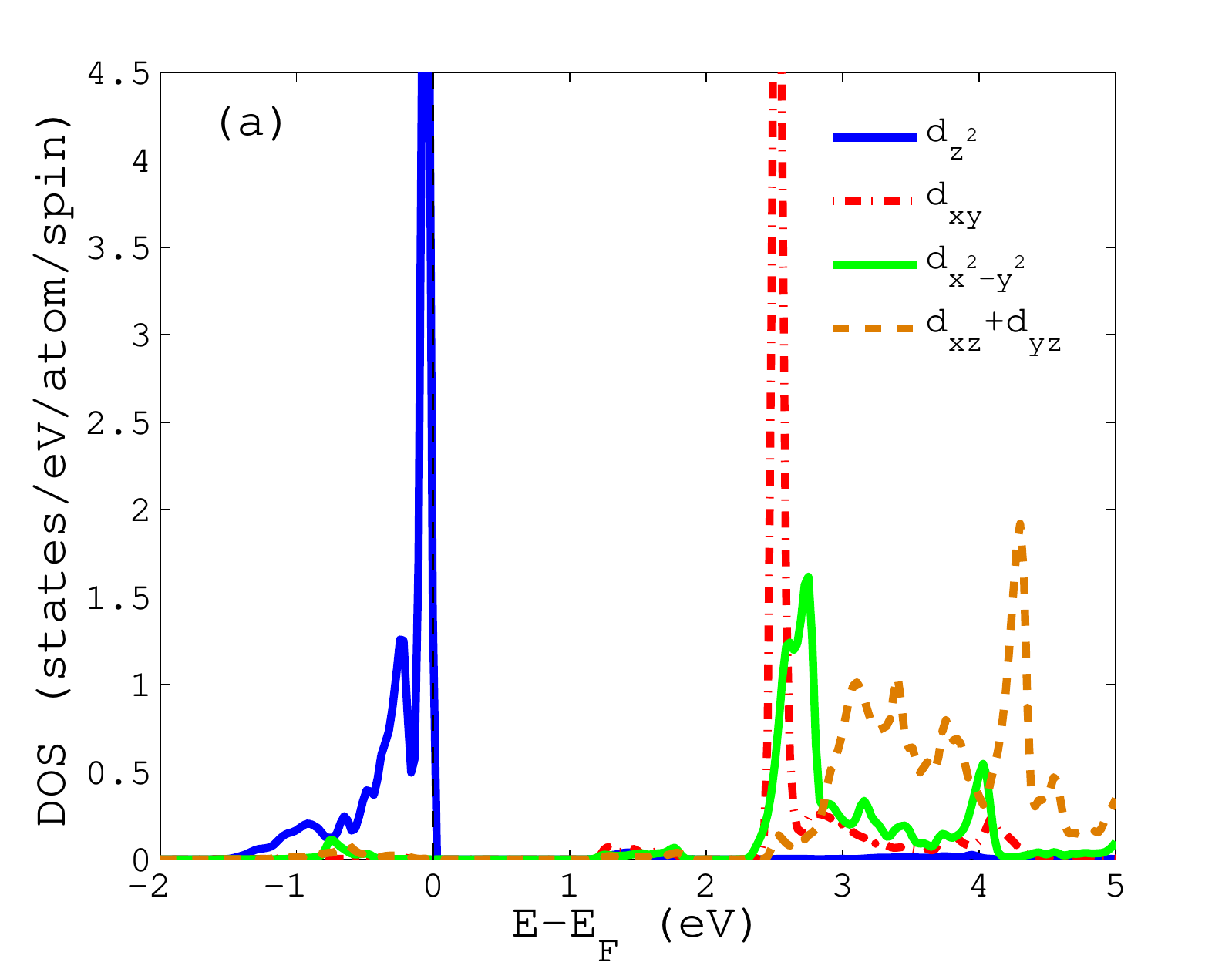}
\includegraphics[width=8cm]{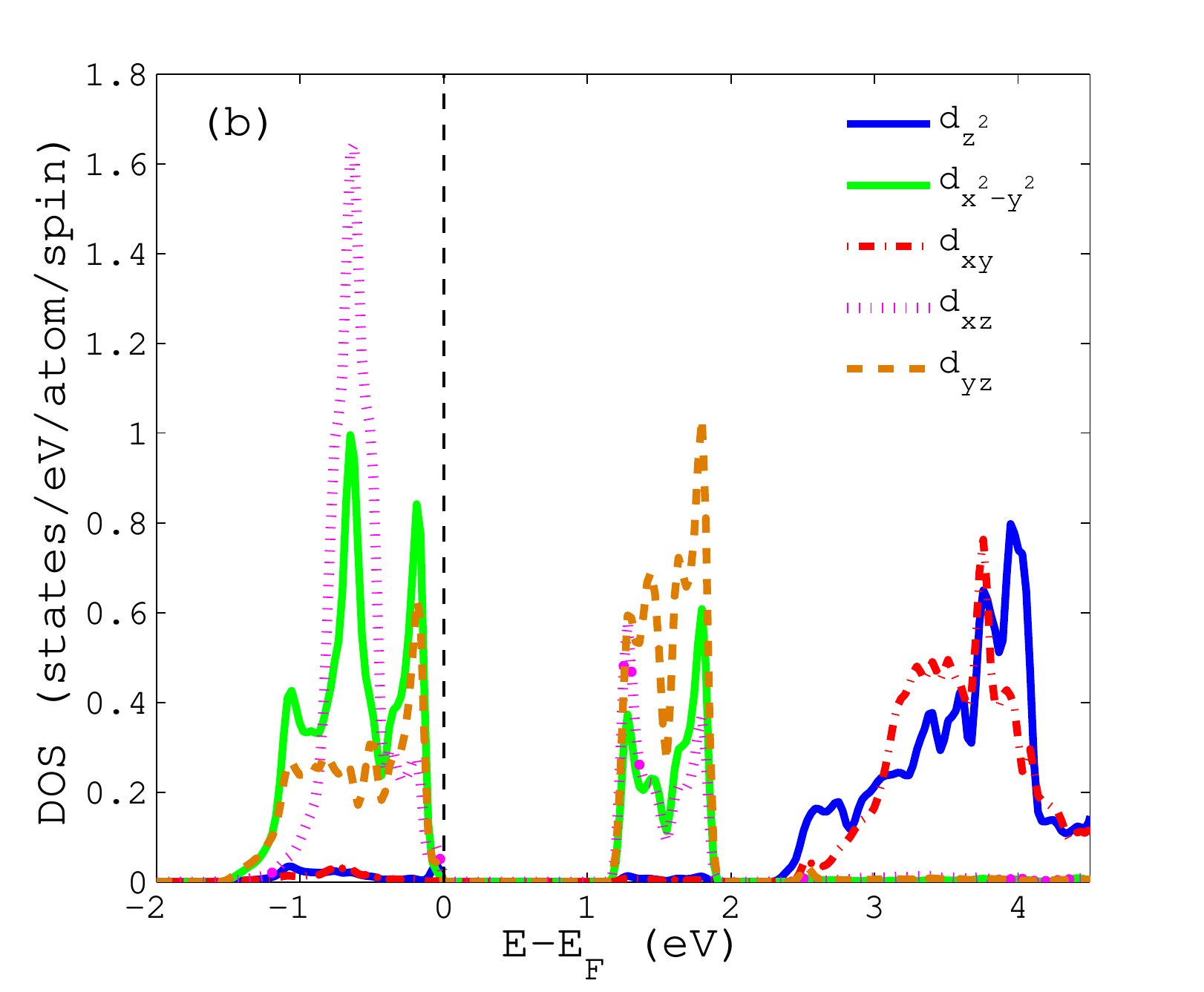}
\caption{\label{dos_75}Spin polarized DOS are shown for x=0.75. (a) Partial DOS of Fe in minority spin channel (b) partial DOS of V in majority spin channel.}
\end{center}
\end{figure}

In experiments it has been observed that in x$>$0.6 regime, \femn  has I4$_1$/a 
symmetry in the tetragonal phase at low temperature and c/a ratio is less than 
1 unlike x$\le$0.6 case where it has I4$_1$/amd symmetry and c/a ratio above 1. 
Therefore, the local distortions
at FeO$_4$ tetrahedra and VO$_6$ octahedra are very different from that 
observed
in x$<$0.6 cases discussed in the previous section (refer to x=0.75 case in 
Table I). 
We observe that in these cases both FeO$_4$ 
tetrahedra as well as VO$_6$ octahedra are compressed along c-direction.
In addition, in the VO$_6$ octahedra four V-O bonds of the ab-plane are split
into two short and two long bonds as opposed to the cases below x=0.6 where
all four V-O bonds in the ab-plane had equal magnitude. These long/short bonds
alternately point along x or y-direction (refer to the Fig 2(a)) in 
successive ab-planes as one moves along c(z)-direction. Trigonal distortions
are also present as listed in Table I.

\begin{figure}
\vspace{-1.0cm}
\begin{center}
\includegraphics[width=9.2cm]{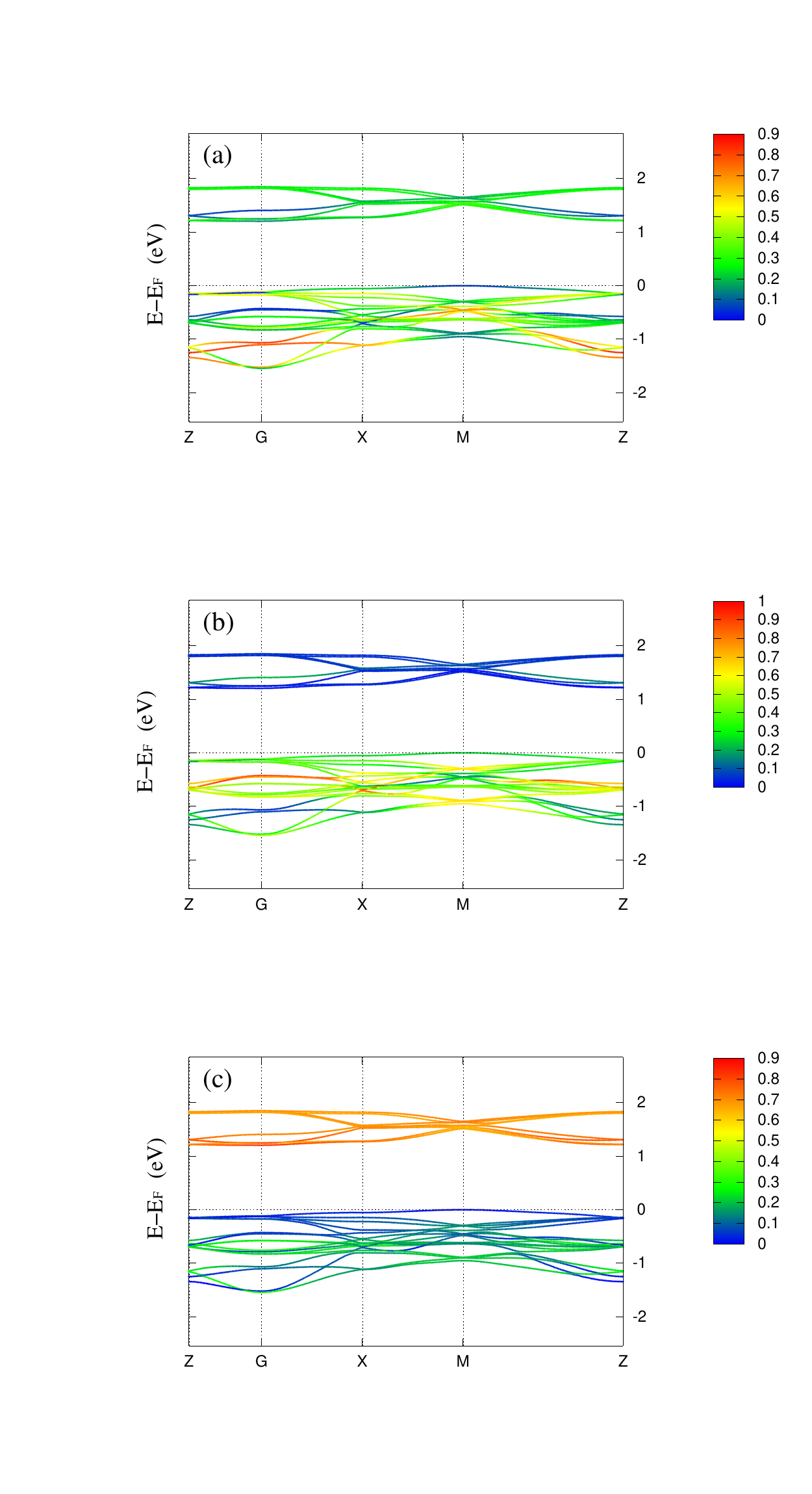}
\caption{\label{wan_proj_75}(Colour online) Projections of V t$_{2g}$ bands on to d$_{xy}$, d$_{yz}$ and d$_{xz}$ Wannier orbitals are shown for x=0.75 in (a), (b) and (c) respectively.}
\end{center}
\end{figure}

We performed electronic structure calculations for \femn at x=0.75 and 1
within GGA+U and GGA+U+SO approximations as we discuss below. As the 
electronic structure and the orbital order (at V sites) are found to be similar in nature for x=0.75 and x=1 we present the results for x=0.75 case.

{\bf GGA+U Results:} In Fig.5 we show the partial DOS for Fe and V d-states in minority
and majority spin channels respectively within GGA+U with U$_{eff}$=4eV and 3eV at Fe(Mn) and V sites.
By looking at the partial DOS of minority Fe d-states we see that JT distortion
in this case makes d$_{z^2}$ the ground state. This is consistent with
the compression of FeO$_4$ tehedra along c(z)-direction. However, the Vanadium
d-states undergo a complicated splitting with all t$_{2g}$ orbital (i.e.
d$_{xy}$, d$_{xz}$, d$_{yz}$) characters being present above and below FL.
This is understandable as VO$_6$ octahedra undergo local and co-operative
JT distortions (both tetragonal and trigonal) leading to orbital ordering. Similar to the previous cases
of x$\le$0.6, we have analysed the orbital characters by calculating the 
corresponding Wannier orbital projections of the V t$_{2g}$-bands above and 
below FL. We show these projections in Fig. 6.
Consistent with the JT compression along c-direction, the lowest bands have
high contribution from d$_{xy}$ orbital. However, in a significant deviation 
from the earlier case (i.e. x$\le$0.6), here we observe that d$_{xz}$ and 
d$_{yz}$ are not equally populated. The filled states below FL have dominant 
contribution from d$_{yz}$ orbital (about 80$\%$ or so) whereas d$_{xz}$'s
presence is very small (about 20$\%$ or so). This implies that the lowest 
occupied
orbital is predominantly d$_{xy}$ in nature whereas the next higher occupied
orbital is d$_{yz}$ in nature as also observed in case of MnV$_2$O$_4$ 
\cite{TM1}. The polarization of the second highest occupied
orbital (i.e. lifting of the degeneracy of d$_{xz}$ and
d$_{yz}$) is consistent with the JT distortion which causes the splitting of V-O
bonds in the ab-plane into two short and two long ones as mentioned above. 
Interestingly, we further observed by plotting the corresponding Wannier
orbitals in real space that the polarization (ordering) character of the
second highest occupied state alternate in successive ab-planes along c-direction. If in one ab-plane it is the d$_{yz}$ which gets
polarized (ordered), then in the next ab-plane it is the d$_{xz}$ that becomes the 
second most populated orbital. 
This keeps repeating so that
the second d-electron of V$^{3+}$ occupies alternately d$_{xz}$ and d$_{yz}$
as we move along c-direction. 
Note that d$_{xy}$ remains
the most populated orbital at all V sites. 
This is often referred to as A-type orbital
ordering in the literature\cite{Garlea,Kawa,DC}. The staggered ordering
of d$_{xz}$ and d$_{yz}$ orbital along c(z)-direction can be understood
if we look at the JT distortion which is also also staggered along the same
direction as described above. If two longer V-O bonds are pointed along x-direction (refer to Fig. 2(a)) then the second highest occupied orbital has dominant
d$_{xz}$ character and if two longer V-O bonds are pointed along y-direction then the second highest occupied orbital has dominant d$_{yz}$ character.  
\begin{figure}
\begin{center}
\includegraphics[width=9cm]{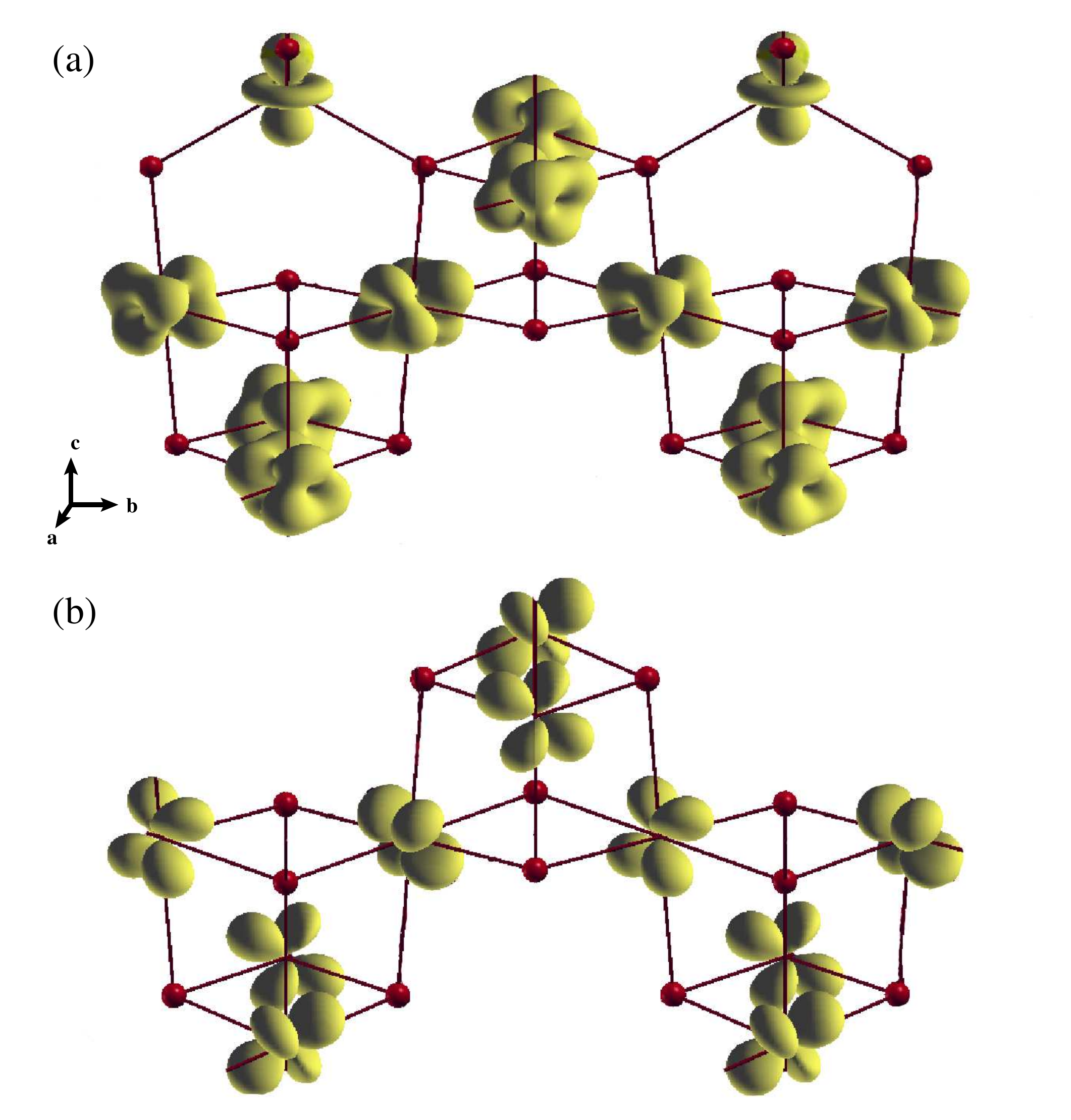}
\caption{\label{el_den_75} 3D electron density is shown in real space lattice
for x=0.75 case in (a). Corresponding hole densities are shown in (b) to depict
the nature of unoccupied orbitals.}
\end{center}
\end{figure}

In Fig.7 we present the 3D electron density in real space for x=0.75 of \femn. 
We clearly see the d$_z^2$ orbitals at Fe sites. The electron density at
V sites are similar to that obtained for the parent compound MnV$_2$O$_4$.
There is a tilting of occupied orbitals within each V chain in
the same ab-plane which was also seen in MnV$_2$O$_4$\cite{TM1}. This tilting
of orbitals along the V chains is caused by the trigonal distortion present.
However, the orbital ordering remains same as we go along the V chains either
in a or b direction. Therefore,
the orbital ordering at V sites as discussed above is applicable to both 
x=0.75 and x=1. We have verified this through Wannier orbital projection 
analysis as presented above for x=1 as well.
\begin{table}
\centering
\begin{tabular}{|p{2.5cm}|p{2.5cm}|p{2.5cm}|}
 \hline
   & Fe $(\mu_B)$ & V $(\mu_B)$ \\
 \hline
 x=0.5 & 0.11 & 0.22\\
 \hline
 x=0.75 & 0.00 & 0.07\\
 \hline
\end{tabular}
\caption{Calculated orbital moments of Fe and V ions in doped systems.}
\label{Tab:3}
\end{table}

{\bf GGA+U+SO Results:} We also performed calculations including SO interaction to study its 
effect on orbital ordering and whether there are complex orbitals involved
in the ordering process. However, we did not observe any significant effect of
SO interaction in this case either.
The calculated orbital moments are also found to be very small as listed in Table II. V orbital
moment is found to be about 0.07$\mu_B$ which is much smaller than that found
for x=0.5 case (i.e. 0.22$\mu_B$). This is consistent with the observation
of Kawaguchi et al. that above x=0.6 the orbital moments should decrease. 
The observation of low orbital moment is also consistent with the recent 
XMCD measurements and previous theoretical calculations on MnV$_2$O$_4$\cite{Nii2,TM1}.
Therefore, we conclude that the orbital ordering involves only real orbitals
in this case as well.

\section{Conclusions}
We have performed first principles density functional theory calculations
in a series of compounds \femn with x = 0, 0.25, 0.5, 0.75 and 1 where 
experimental measurements related to orbital ordering have recently been 
reported. We investigated
in detail the orbital ordering of Fe and V d-orbitals in these systems
from our calculations under various approximations such GGA, GGA+U and 
GGA+U+SO. 
We have also performed a detail analysis of long range orbital order by
calculating the Wannier orbital projections of the occupied d-bands and 
plotting those orbitals in real space lattice. On the basis of the 
observed orbital order from our calculations we divide these systems into 
two classes; one which includes the compounds x=0,0.25 and 0.5 and the other
includes x=0.75 and 1. In the former, we observe d$_{x^2-y^2}$ orbital order at Fe sites while in the later it is d$_{z^2}$ orbital ordering. 
As far as V sites are concerned, in the former case, d$_{xy}$ orbital is found 
to be present at all V sites as expected and the second orbital that is ordered has equal contribution
from d$_{xz}$ and d$_{yz}$ (i.e a linear superposition of the type d$_{xz}$+d$_{yz}$ or 
d$_{xz}$-d$_{yz}$). In successive ab-planes along c-direction 
the composition of this orbital alternates between d$_{xz}$+d$_{yz}$ and 
d$_{xz}$-d$_{yz}$. We further observe that real orbitals, not complex ones 
as had been claimed in some reports earlier\cite{Kawa}, are involved
in the orbital ordering process. This observation is consistent with the recent
XMCD measurements on the parent compounds\cite{Nii2}.  For x=0.75 and 1, we observe, apart from d$_{xy}$ orbital being  
present at all V sites, the second electron occupies preferentially
either d$_{xz}$ or d$_{yz}$ in successive ab-planes as one moves along 
c-direction which is often referred to as A-type orbital ordering in the literature. 

\section{Acknowledgments}
Authors acknowledge D. Choudhuri for useful discussions particularly on experimental observations in these systems. DD acknowledges DST Inspire (India) for research fellowship. TM acknowledges CSIR (India) for funding under the project grant no: 03(1212)/12/EMR-II.

\end{document}